\documentclass[oldversion]{aa}

\usepackage{natbib}
\bibpunct{(}{)}{;}{a}{}{,}
\usepackage{graphicx}
\usepackage{amssymb}

\title{On the intensity interferometry and the second-order correlation function $g^{(2)}$ in astrophysics}

\author{C. Foellmi\inst{1}}
\institute{Laboratoire d'Astrophysique de Grenoble, Universit\'e Joseph Fourier, CNRS UMR5571, 414 rue de la Piscine, 38400 Saint-Martin d'H\`eres, France.}
\offprints{C. Foellmi \email{cedric.foellmi@obs.ujf-grenoble.fr}}

\abstract{Most observational techniques in astronomy can be understood as exploiting the various forms of the first-order correlation function $g^{(1)}$. As however demonstrated by the Narrabri Stellar Intensity Interferometer back in the 1960's by Hanbury Brown \& Twiss, and which is the first experiment to measure the second-order correlation function $g^{(2)}$, light can carry more information than simply its intensity, spectrum and polarization. Since this experiment, theoretical and laboratory studies of non-classical properties of light have become a very active field of research, namely quantum optics. Despite the variety of results in this field, astrophysics remained focused essentially on first-order coherence. In this paper, we study the possibility that quantum properties of light could be observed in cosmic sources. We provide the basic mathematical ingredients about the first and the second order correlation functions, applied to the modern context of astronomical observations. We aim at replacing the Hanbury Brown \& Twiss experiment in this context, and present two fundamental limitations of an intensity interferometer: the requirement of a chaotic light source, and the rapid decreases of the amount of correlated fluctuations with surface's temperature. The first of these limitations emphasize paradoxically the fact that the exploitation of $g^{(2)}$ is richer than what a modern intensity interferometer could bring and is particularly interesting for sources of non-thermal light. We also discuss new photon-counting avalanche photodiodes currently being developed in Grenoble, and their the impact on limiting magnitudes of an intensity interferometer. We conclude by briefly presenting  why microquasars in our galaxy and their extragalactic parents can represent an excellent first target in the optical/near-infrared where to observe non-thermal light, and test the use of \gtwo\ in astrophysical sources.}


\date{Received $<$date$>$, Accepted $<$date$>$}

\titlerunning{Intensity Interferometry and the second-order correlation function in astrophysics.}

\newcommand{\gone}{$g^{(1)}$}
\newcommand{\gtwo}{$g^{(2)}$}

\begin{document}
\maketitle

\section{Introduction}

There is nowadays a revival of interest in the literature about the intensity interferometry\footnote{An interest strong enough for the International Astronomical Union to create a dedicated Working Group.} \citep{LeBohec-Holder-2005,Ofir-Ribak-2006a,Ofir-Ribak-2006b,Dravins-2008,LeBohec-etal-2008,deWit-etal-2008a,deWit-etal-2008b}\citep[see also, for instance,][on specific applications of intensity correlations]{Borra-2008,Jain-Ralston-2008}. Intensity interferometry differs from the phase interferometry technique (also called amplitude interferometry) as it measures correlations between light intensities instead of the interferences of electromagnetical fields. These correlations can also be used to measure stellar radii as shown by the Narrabri Stellar Intensity Interferometer (hereafter NSII) installed and operated by Prof. Hanbury Brown between 1962 and 1972 in Narrabri in Australia \citep{HanburyBrown-1974}. This experiment has a peculiar role in astrophysics. It was the first to measure the second-order correlation function $g^{(2)}$. As such it opened the door to the observation of phenomena that cannot be explained classically, but requiring a full quantum mechanical treatment. Since then, the so-called Hanbury-Brown \& Twiss (HBT) effect is used nowadays in various fields of modern physics \citep[see for instance][]{Baym-1998,Alexander-2003}, but not anymore in astrophysics; the main reason being its poor sensitivity.

Hanbury Brown \& Twiss explained mathematically the equivalence between a classical and a quantum analysis of the effect they observed \citep{HanburyBrown-Twiss-1957a,HanburyBrown-Twiss-1958a, HanburyBrown-Twiss-1958b, HanburyBrown-Twiss-1958c}. They emphasized that the observed phenomenon exemplified the wave rather than the particle aspect of light, and claim originally that it has no dependence on the actual mechanism by which the light energy was originally generated. This is not true and this important point is discussed below.

Technical aspects aside, Hanbury Brown \& Twiss succeeded at measuring stellar radii with their Intensity Interferometer because of the following fundamental relationship:
\begin{equation}
	\label{g1g2}
	\left| g^{(1)} \right|^2 = g^{(2)} - 1
\end{equation}
where \gone\ and \gtwo\ are the first- and second-order correlation functions respectively\footnote{Parenthesis indicate that the number is not an exponent.}. The modulus of \gone\ corresponds exactly to the definition of the usual {\it visibility} measured by a Michelson interferometer. The hidden condition for this equation to be valid is that the light produced by the source must be chaotic. Even if this may be true for a vast majority of astrophysical sources, it allows to draw the boundaries of validity of the technique.

The possible exploitation of \gtwo\ in astrophysics is probably far more richer than simply a modern intensity interferometer, as shown by the experiments elsewhere in physics. Cosmic light sources where there are no intensity fluctuations, or where the light emission is not (fully) chaotic could therefore potentially represent astronomical versions of non-classical light sources created in the laboratory. In this case, Equ.~\ref{g1g2} is not valid and \gtwo\ will carry additional information that is not contained in \gone. This paper aims to be a first step to see if meaningful astrophysical measurements on such sources can be achieved not only to foster progresses on cosmic objects but also on light itself. 

In this paper, we review the fundamental meaning of the relationships between \gone\ and \gtwo. We aim at giving the basic framework to understand why most current techniques in astrophysics exploit only \gone, and what means a transition towards the use of \gtwo, not only through an intensity interferometer (hereafter II), but also for new and so far unaccessible observables. Along the way, the intensity interferometer of Narrabri can be put in context. We discuss the fundamental limitations of an II, and and evaluate them against the capabilities of new avalanche photodiodes currently being developed in Grenoble. We finish by discussing some speculative ideas to go beyond the stellar II and exploit the second-order correlation function in astrophysics.

\section{Mathematical basics}

\subsection{The first-order correlation function}

The definition of the normalized first-order correlation function reads:
\begin{equation}
\label{equ_g1_def_full}
g^{(1)}(\vec{r_1}, t_1, \vec{r_2}, t_2) = \frac{ \langle E^*(\vec{r_1}, t_1)E(\vec{r_2}, t_2) \rangle }{ \left[ \langle \left| E(\vec{r_1}, t_1)\right|^2 \rangle \langle \left| E(\vec{r_2}, t_2)\right|^2 \rangle \right]^{1/2} }
\end{equation}
where $E(\vec{r}, t)$ is the total complex amplitude of the electromagnetical field of the beam light. A measurement of \gone\ evaluates the first-order coherence which is equal to the statistical average (denoted by angle brackets) of the correlation between electromagnetical fields. Considering only the temporal coherence which depends on the time difference variable $\tau = t_2-t_1$, we have:
\begin{equation}
\label{equ_g1_def}
g^{(1)}(\tau) = \frac{\langle E^{*}(t) E(t+\tau) \rangle}{\langle E^*(t)E(t) \rangle}
\end{equation}
Note that in the case of a plane-parallel wave where only the direction parallel to the beam can be considered, $\tau$ might be redefined as $\tau = t_2-t_1 + (z_2-z_1)/c$.

\begin{figure*}
\centering
\includegraphics[width=0.45\linewidth]{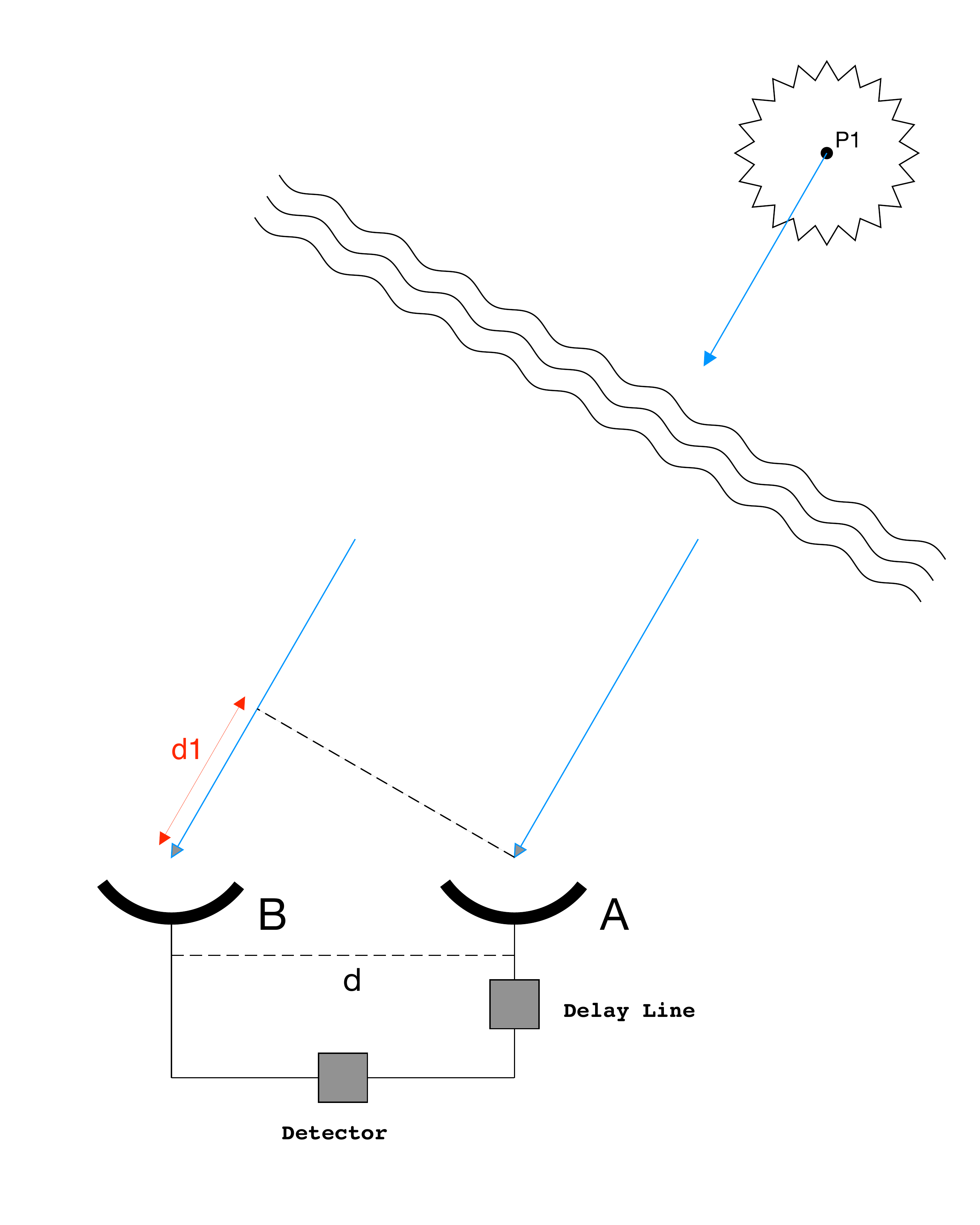}
\includegraphics[width=0.45\linewidth]{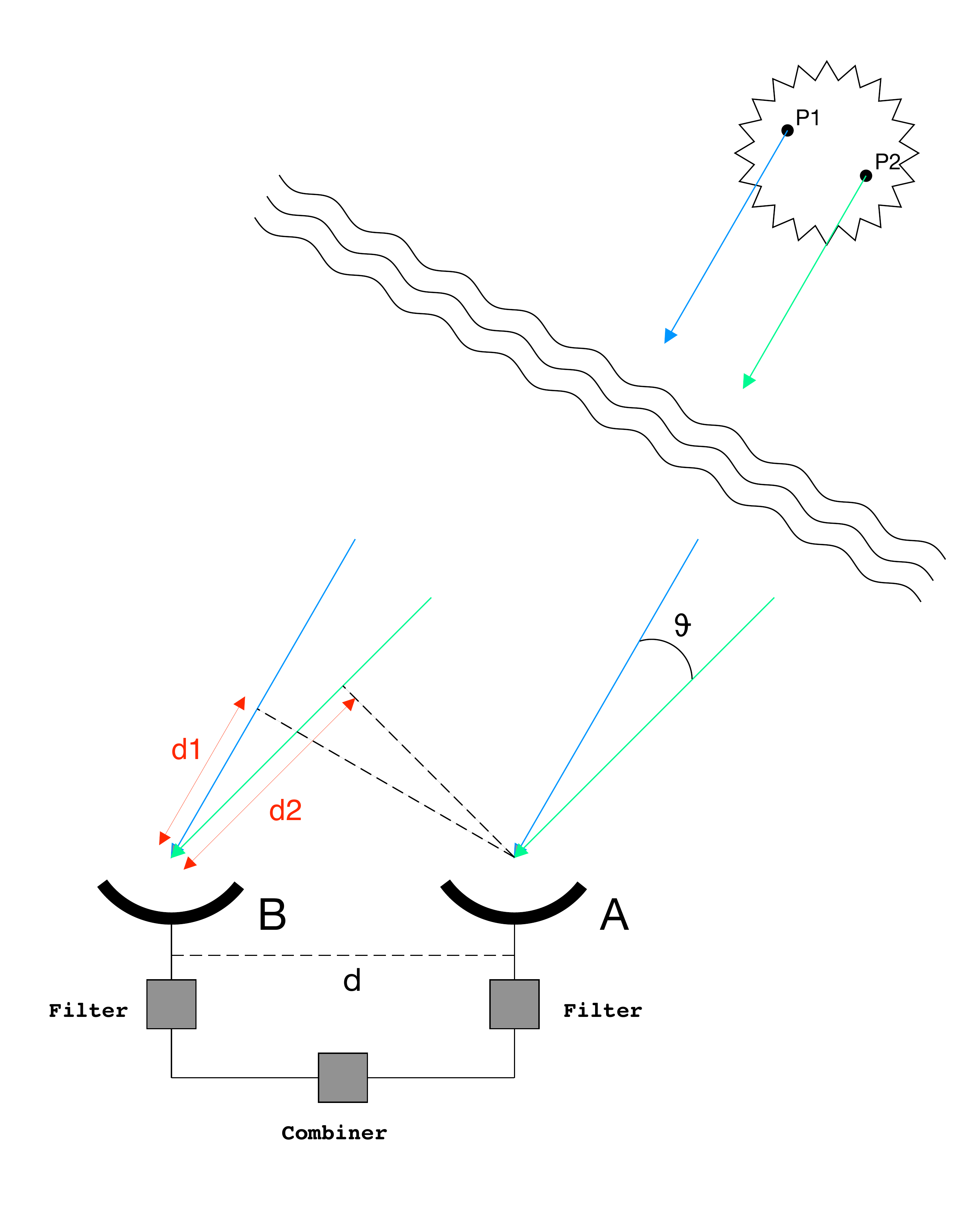}
\caption{Working principle of an intensity interferometer compared to that of a Michelson interferometer. On the left (Michelson interferometer), a first-order spatial coherence is made, associated with the statistical average of single point sources. Delay lines are used to cancel out the path difference $d_1$ and ensure temporal coherence. This cancellation must be achieved to a precision that has to be smaller than the coherence length of the light at the given wavelength: $\Delta x \lesssim c/\tau_c$. On the right (intensity interferometer), a second-order spatial coherence is made, associated with the statistical average of the correlations between {\it pairs} of point sources. Path differences do not need to be cancelled to the same precision, since the dominant fraction of this path difference due to azimuthal distance and atmosphere fluctuations is identical for both paths. However, the equality of both paths must be achieved to a level corresponding to the difference of coherence time between the two Fourier components: $\Delta x \lesssim c/(\tau_{c1}-\tau_{c2}) = \frac{c}{\tau_{c1}(1-\tau_{c2}/\tau_{c1})} \gg c/\tau_{c1}$ when $\tau_{c1} \sim \tau_{c2}$. This explains the less stringent need in high-precision optical devices and the very poor sensitivity of the II against seeing fluctuations.}
\label{fig_ii_principle}
\end{figure*}

The case of a Michelson interferometer is schematized in Fig.~\ref{fig_ii_principle} (left panel). Stellar radii are measured through the evaluation of the spatial coherence of the light, while $t_1 = t_2$ is ensured by delay lines. The evaluation of the spatial coherence of the light is a measurement of how far two points sources can lie in a place transverse to the propagation direction of the light, and still be correlated in phase.

The path difference $d_1$ in Fig.~\ref{fig_ii_principle} is cancelled out by mechanical means to ensure that the evaluation of the electromagnetical wave at two different locations remains {\it temporally} coherent. Somehow, the role of the chaotic light is already present in here: if the electromagnetic wave was temporally first-order coherent for any time difference $\tau$, we would not need to equalize the light paths. Another possibility to get closer to this situation is to place a high-resolution dispersive element which will render each wavelength channel closer to a monochromatic source.

In the case of chaotic light and broad-band observations, the temporal coherence time is very short. Hence the fringes vanish rapidly with a time difference $\tau \equiv t_2 - t_1$ larger than the light coherence time $\tau_c$, whatever the degree of spatial coherence. We can say that our ability to see the variation of the spatial coherence when changing the distance $d$ between the two telescopes depends on the condition that the temporal coherence is fulfilled.

By definition, the visibility, also called fringe contrast, is the modulus of \gone:
\begin{equation}
V = \left| g^{(1)}(\vec{r_1}, t_1, \vec{r_2}, t_2) \right|
\end{equation}

In some sense, a (spatial) visibility can be considered as the most basic element of an image. It is directly proportional to the Fourier Transform (FT) of the angular distribution of the light intensity on sky (along the projected telescopes baseline), according to the van Cittert-Zernike theorem \citep[see e.g.][]{Haniff-2007a}. Although unusual, any imaging device can actually be considered as a interferometric instrument performing the FT operation and transforming the plane wavefront into intensity peaks corresponding to the source positions. Similarly, one can invoke the Wiener-Khintchine theorem to reconstruct the spectral density distribution of a source by measuring the temporal coherence of the light wave (see below). The interferences are either directly recorded (like in a Fourier Transform spectrograph), or a dispersive element (like a grating) performs the FT operation and the spectrum is directly recorded onto the detector.

Roughly, one can say that a visibility, a spectrum or an image of a celestial source are the different forms of the evaluation of always the same {\it first-order} function, \gone. In the quantum limit, it means measuring the properties of single photons (to the contrary of collective properties of a photon stream). So far, observational astronomy has not made use of \gtwo\ (or any higher-order correlation function), with the exception of the NSII\footnote{Strictly speaking, the NSII was preceded by two experiments of the same nature, performed in Jodrell Bank (England) under the lead of R. Hanbury Brown. The first one was operating in the radio frequencies and was used to measure the angular diameter of the Sun, and to study Cassiopeia A and Cygnus A. The second one was a crude optical II dedicated to prove the feasibility of the concept, by measuring the angular diameter of Sirius. See \citet{HanburyBrown-1974} for the details and subsequent references. Historically, the NSII remains however the true experiment demonstrating the correlations of intensity fluctuations.}.

\subsection{The nature of the difference between temporal and spatial coherence}

There is an important difference between the information obtained through the measurement of temporal coherence compared to that of the spatial coherence. The former allows to probe the physical processes of light production in single or multiple atom systems, while the second is related to the spatial "properties" of the emitting system itself. As a matter of fact, in the first order, the temporal coherence of light is directly related to the spectral density of the source, according to the Wiener-Khintchine theorem:
\begin{equation}
F(\omega) = \frac{1}{\pi} \Re\left( \int_0^{\infty} g^{(1)}(\tau) \exp(i\omega\tau) \, d\tau \right)
\end{equation}
where the symbol $\Re$ denotes the real-part of a complex number.

Although being an information about the object itself too, a spectrum provides mostly information on the light emission conditions. In other words, the study of light production (in any correlation order) is certainly another mean to foster progresses in our understanding of cosmic {\it objects}, but in a manner which is closer to what high-energy astrophysics do. When one are interested to light production and light properties of higher order or non-classical, it is thus natural, because of the fact that Equ~\ref{g1g2} is a restricting condition, to look for objects where the light emission is not fully disturbed by random processes, like in star's atmospheres.

\subsection{The second-order correlation function}

By definition, the general form of the (temporal) normalized second-order correlation function writes:
\begin{equation}
\label{g2_def}
\label{g2_E}
g^{(2)}(\tau) = \frac{\langle E^*(t)E^*(t+\tau)E(t+\tau)E(t) \rangle}{\langle E^*(t)E(t) \rangle^2}
\end{equation}
An important overlooked point in the above equation is that generally, the complex electromagnetic fields $E$ do not commute. Therefore, in the general case, the order of the terms cannot be rearranged. In the case of chaotic light however the cancellation of cross-terms between random relative phases allows this reorganization, and we can write:
\begin{eqnarray}
\label{g2_I} g^{(2)}(\tau) & = & \frac{\langle E^*(t)E(t)E^*(t+\tau)E(t+\tau)\rangle}{\bar{I}^2} \\
 & = & \frac{\langle \bar{I}(t) \bar{I}(t+\tau) \rangle}{\bar{I}^2} 
\end{eqnarray}

Using Equ.~\ref{g2_def}, we can show that \citep[][Equ.~3.7.13, p.110]{Loudon-2000}, for $n$ independent radiative atoms:
\begin{eqnarray}
\nonumber \langle E^*(t)E^*(t+\tau)E(t+\tau)E(t)\rangle & = & \\
\nonumber && \hspace{-3.5cm} \sum_{i=1}^n \left\langle E_i^*(t)E_i^*(t+\tau)E_i(t+\tau)E_i(t)\right\rangle +  \\
\nonumber && \hspace{-3.5cm} \sum_{i\neq j}\{ \left\langle E_i^*(t)E_j^*(t+\tau)E_j(t+\tau)E_i(t)\right\rangle \\
&& \hspace{-2.5cm} + \left\langle E_i^*(t)E_j^*(t+\tau)E_i(t+\tau)E_j(t) \right\rangle \}
\end{eqnarray}

Only the terms where the field of a given atom is multiplied by its complex conjugates are kept. All other terms vanish because the average of the waves of the different atoms with random relative phases are zero. Therefore:
\begin{eqnarray}
\nonumber \langle E^*(t)E^*(t+\tau)E(t+\tau)E(t)\rangle & = & \\
\nonumber && \hspace{-3.5cm} n \langle E_i^*(t)E_i^*(t+\tau)E_i(t+\tau)E_i(t)\rangle + \\
\nonumber && \hspace{-3.5cm} n(n-1) \left\{ \left\langle E_i^*(t)E_i(t)\right\rangle^2 + \left|\left\langle E_i^*(t)E_i(t+\tau) \right\rangle\right|^2\right\}
\end{eqnarray}
In the usual case where $n \gg 1$, the second term dominates largely. We have therefore:
\begin{eqnarray}
\nonumber \langle E^*(t)E^*(t+\tau)E(t+\tau)E(t)\rangle & = & \\
\nonumber && \hspace{-3cm} n^2 \left\{ \langle E_i^*(t)E_i(t)\rangle^2 + \left|\langle E_i^*(t)E_i(t+\tau)\rangle\right|^2\right\}
\end{eqnarray}

Using the definition of \gone\ and \gtwo, we obtain Equ~\ref{g1g2}:
\begin{equation}
g^{(2)}(\tau) = 1 + \left|g^{(1)}(\tau)\right|^2 \hspace{1cm} \textrm{for } n \gg 1
\end{equation}

Note that \gtwo\ is a real number to the contrary of \gone. To recover the phase information, we can still use the amplitude itself thanks to the analyticity of the discrete Fourier Transform \citep[see for instance][although an image flip degeneracy remains]{Holmes-Belenkii-2004}, or the high-order correlations \citep[see for instance][]{Zhilyaev-2008}, but with the issue of an even lower sensitivity.

The correlator of an intensity interferometer is designed to measure the quantity:
\begin{equation}
\frac{\left\langle \left[ \bar{I}(\vec{r}, t) - \bar{I}\,\right] \left[ \bar{I}(\vec{r}+\vec{\rho}, t) - \bar{I}\,\right] \right\rangle}{\bar{I}^2} = g^{(2)}(\vec{\rho}) - 1
\end{equation}
where $\vec{\rho}$ indicates the distance vector between the two telescopes, as in Fig.~\ref{fig_ii_principle}, and the mean light intensities $\bar{I}$ and $\bar{I}(\vec{r}, t)$, are defined as:
\begin{eqnarray}
\bar{I} & = & \langle \bar{I}(t) \rangle \\
\bar{I}(t) & = & \frac{1}{2}c \, \epsilon_0 \left|E(t)\right|^2
\end{eqnarray}
where it is assumed that the light source is stationary (i.e. its statistical properties does not change with time).

\subsection{The demonstration by Hanbury Brown of the working principle of the intensity interferometer}

In order to clarify what an intensity interferometer measures, and how it is related to the above formalism, let us briefly review the explanation provided by Hanbury Brown \citep[][Chap.~2]{HanburyBrown-1974}. The basic principle is schematized in Fig.~\ref{fig_ii_principle} (right panel). Intensity interferometry is a second-order measurement, and instead of evaluating the electromagnetic fields in two different locations, it evaluates the {\it correlations} between {\it pairs} of point sources. Each points radiate white light and is completely independent of one another.

A lightwave can be decomposed as a superposition of a large number of sinusoidal components, each component having a steady amplitude and phase over the period of observation but both the amplitude and phase being random with respect to the other components. Let us consider only one component of this superposition (Hanbury Brown calls it a Fourier component) reaching telescope $A$ from $P_1$ and another component with different frequency reaching the same telescope $A$ from the second point $P_2$. We have:
\begin{eqnarray}
C_1 = E_1 \sin(\omega_1 t + \phi_1) \\
C_2 = E_2 \sin(\omega_2 t + \phi_2)
\end{eqnarray}
where $E$ denotes the electromagnetic field amplitude of the given component.

The intensities are transformed into electrical currents in the photodetectors. The output current as measured behind telescope $A$ is proportional to the {\it intensity} of the light. Assuming linear polarization:
\begin{equation}
I_A = K_A \left[ E_1 \sin(\omega_1 t + \phi_1) + E_2 \sin(\omega_2 t + \phi_2)\right]^2
\end{equation}
where $K_A$ is a constant of the detector. The {\it same} Fourier component will also reach telescope $B$, with path differences $d_1$ and $d_2$. Hence:
\begin{eqnarray}
I_B & = & K_B [ E_1 \sin(\omega_1 (t + d_1/c) + \phi_1) + \\ 
\nonumber    &   & \hspace{2cm} E_2 \sin(\omega_2 (t + d_2/c) + \phi_2)]^2
\end{eqnarray}

If we expand the above two equations, we obtain each time 4 terms. The first two terms are the sum of the light intensities from both components, and the third corresponds to the intensity of the sum of frequencies ($\omega_1 + \omega_2$). All these terms can be filtered out, and we are left with the only term of interest here, corresponding to the difference of the frequencies ($\omega_1 - \omega_2$):
\begin{eqnarray}
I_A & = & K_A E_1 E_2 \cos[(\omega_1-\omega_2)t + (\phi_1-\phi_2)] \\
\nonumber I_B & = & K_B E_1 E_2 \cos[(\omega_1-\omega_2)t + (\phi_1-\phi_2) + \\
& & \hspace{4cm} (\omega_1 d_1 - \omega_2 d_2)/c]
\end{eqnarray}

These two components are correlated because they have the same beat frequency $(\omega_1 -\omega_2)$ and differ only in phase by the quantity $(\omega_1 d_1 - \omega_2 d_2)/c$. If we put for simplicity $\omega_1 \approx \omega_2 = \omega$, the correlation, which is the product of $I_A$ and $I_B$ writes:
\begin{equation}
c(d) = K_A K_B E_1^2 E_2^2 \cos\left[\frac{\omega}{c}(d_1-d_2)\right]
\label{ii_correlation_final}
\end{equation}

The phase difference between these correlated components is not the phase difference of the light waves at the two detectors but is the {\it difference} between the phase differences (or the difference between the relative phases observed independently at the two telescopes) of the two Fourier components. By simple geometry, and because $\theta$ is a very small angle ($\sin(\theta) \approx \theta$), we have:
\begin{equation}
c(d) = K_A K_B E_1^2 E_2^2 \cos\left[2\pi d \theta / \lambda\right]
\end{equation}
where $d$ is the separation between the two telescopes. Finally \citet{HanburyBrown-Twiss-1957a} have shown that when computing the total correlation by integrating Equ.~\ref{ii_correlation_final} over all possible pairs of points on the disk of the star, over all possible pairs of Fourier components (which lie within the optical bandpass) and over all beat frequencies (which lie within the bandpass of the electrical filters) we obtain that the correlation is proportional to the square of the modulus of the visibility.

Clearly, by no means in this analysis the hypothesis of chaotic light is explicitly made. It plays nonetheless an important role.

\subsection{The implicit hypothesis of the chaotic nature of light in the Narrabri Stellar Intensity Interferometer experiment}

As said in the introduction, \citet{HanburyBrown-Twiss-1957a} never made the hypothesis that the light source must be producing chaotic light. Note that we use here the term "chaotic" instead of "thermal", since chaotic light can be produced by independent non-thermal individual sources. The subtle distinction has its importance in the case of radio masers, as discussed below.

It is interesting to note that the year 1957 is also the time of the development of lasers \citep[e.g.][]{Schawlow-Townes-1958}, the properties of which were probably not fully known and understood at that time. In their paper of 1957, Hanbury Brown and Twiss even claim that the effect of correlation between intensity fluctuations had nothing to do with the mechanism by which the light was actually produced. They end their paper by saying: "[...] still less does it imply that the photons must have been injected coherently into the radiation field. On the contrary, if one wishes to picture the electromagnetic field as a stream a photons, one has to imagine that the light quanta redistribute themselves over the wavefront, as the radiation field, which may be quite incoherent in origin, is focused and collimated into beams capable of mutual interference; thus the correlation between photons is determined solely by the energy distribution and coherence of the light reaching the photon detectors." 

Interestingly, Hanbury Brown made no reference to the nature of light itself when presenting the above analysis based on Fourier components, noting that this way is "freer from conceptual traps". However, in his book in 1974 he presented the analysis in a slightly different manner, although not emphasizing the radical difference it implies for the interpretation of the phenomenon. At the beginning of Chap.~3, he did start by stating that "white light of the thermal origin has the properties of a Gaussian random process" as shown by several classical papers. This property is being used further in his analysis but never to express how the results could have been different in the case of non-thermal light. In the formalism presented above, it lies precisely at the point where we identify Equ.~\ref{g2_E} to Equ.~\ref{g2_I}.

It is known however today that photons may follow different statistics and that it depends precisely on how the collections of photons are emitted. This does not invalidate the analysis presented above, but it does prevent to interpret the correlation seen in \gtwo\ as a measurement of the visibility $\left|g^{(1)}\right|$, and hence as an information about the angular distribution of light intensity along a baseline separating two telescopes via the classical van Cittert-Zernike theorem. In other words, Equ.~\ref{g1g2} is not satisfied for non-chaotic light. For instance, it can be shown that, for an ideal monochromatic linearly-polarized laser, we have $g^{(1)}(\tau) = g^{(2)}(\tau) = 1$, independent of $\tau$. It also means that no additional information at all is encoded in higher-order functions in chaotic light. In fact, one can show that for such type of light only, it is always possible to write the $n^{th}$ order of the correlation function as sums of products of the first order function \citep[][p. 115]{Glauber-2007}. 

\subsection{Photon statistics and non-classical light}

Interestingly, different types of light can be classified thanks to \gtwo$(\tau$), or more precisely, depending on their value of \gtwo$(\tau\sim0)$.

\begin{figure}
\centering
\includegraphics[width=0.9\linewidth]{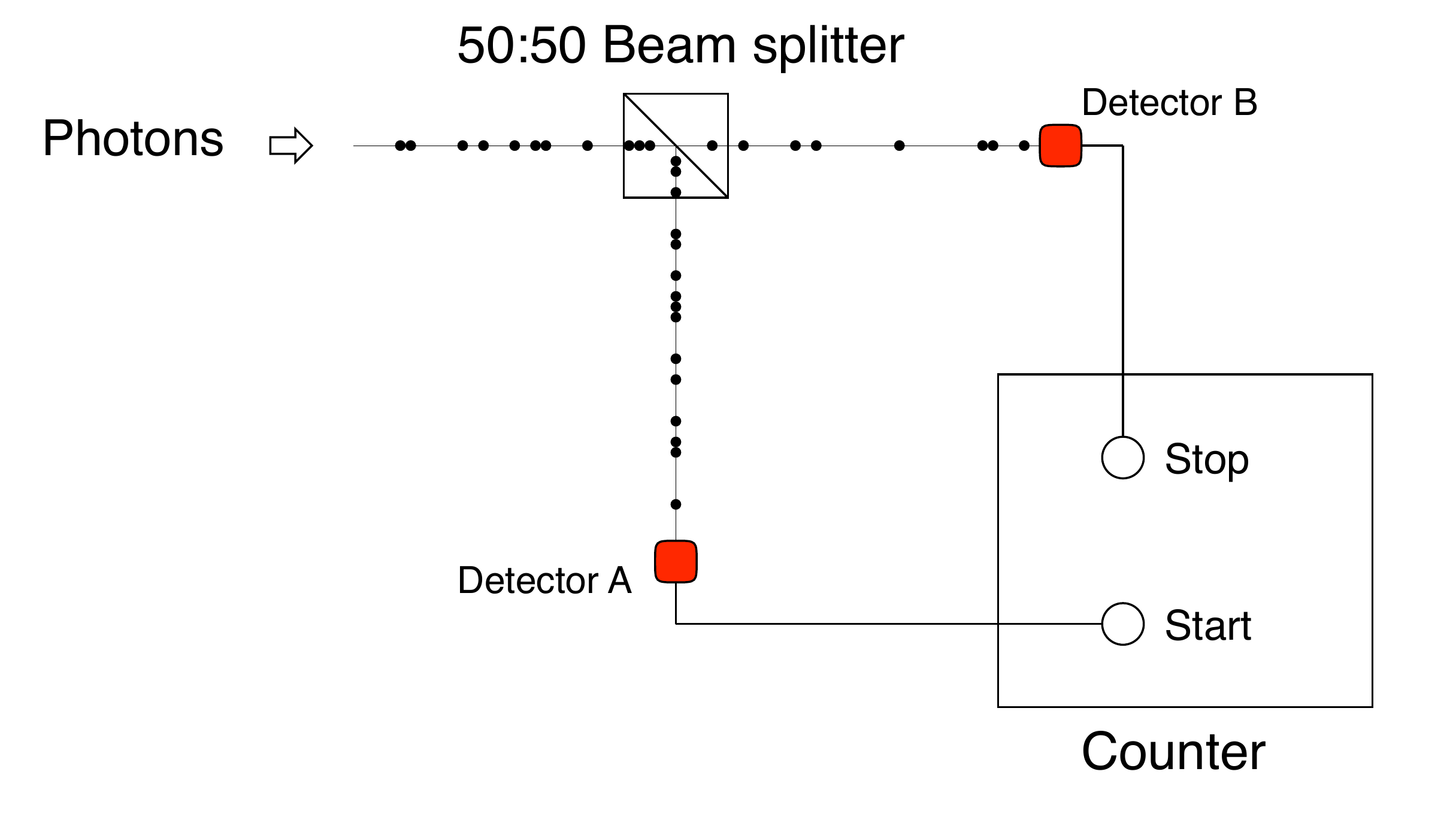}
\caption{Sketch of a typical experimental setup to measure \gtwo. Light or photons come from the left and enter the 50:50 beamsplitter. If a photon is detected on detector A, a counter is started. This counter runs until another photon is recorded in Detector B \citep[see][]{Fox-2006}. It corresponds basically to the Narrabri interferometer setup, if the beam splitter is replaced by two telescopes recording an intensity each. The principle remains however the same.}
\label{fig_HBT}
\end{figure}

\begin{figure}
	\centering
	\includegraphics[width=0.8\linewidth]{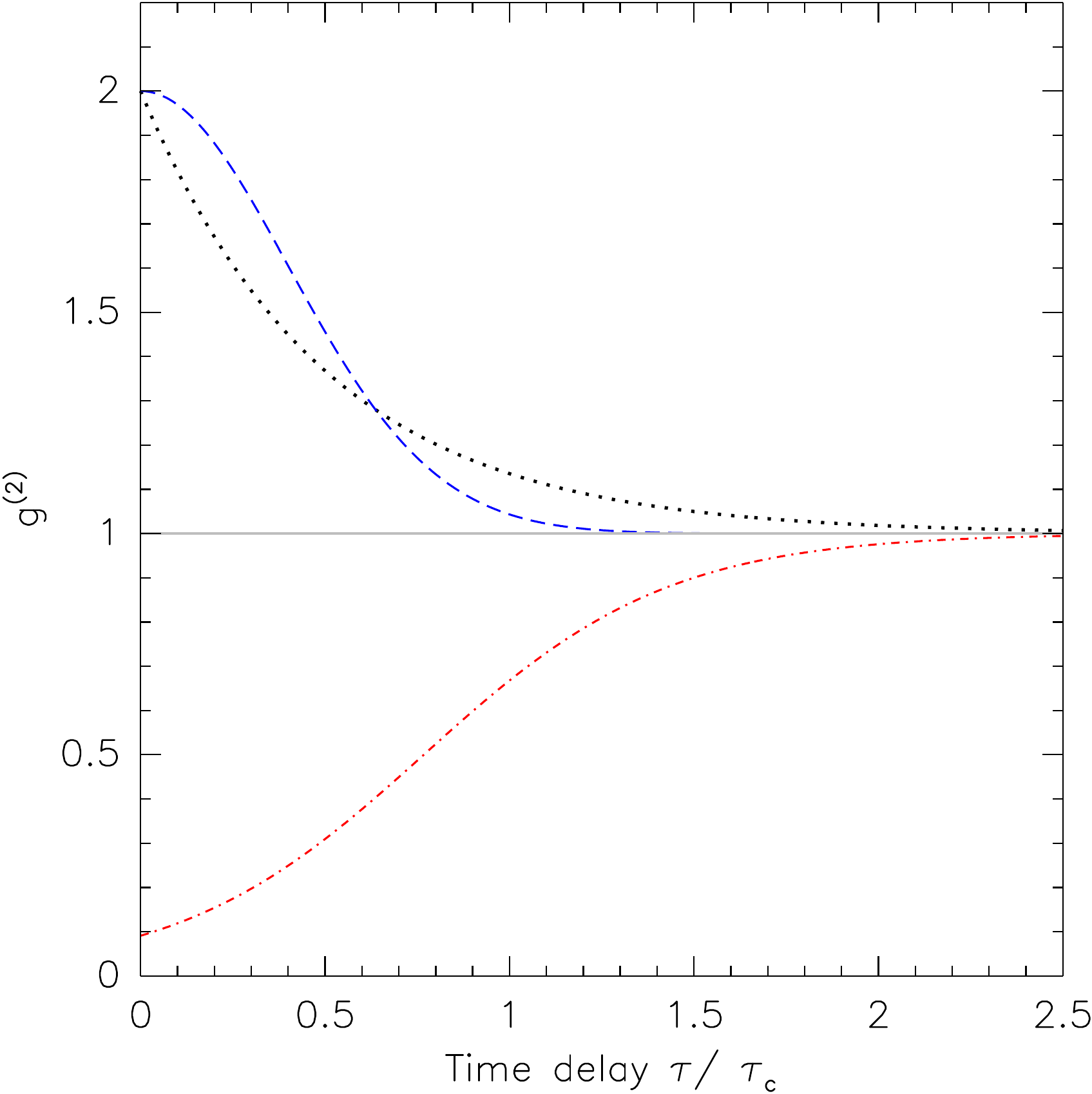}
	\caption{Example of \gtwo\ curves for light of coherence time $\tau_c$. In blue dashed a black-body light with gaussian broadening, in black dotted with Lorentzian broadening. In  red dot-dashed is shown an example of \gtwo\ curve for anti-bunched light whose signature is \gtwo$(0) < 1$.}
	\label{fig_g2}
\end{figure}

A typical experimental setup to measure \gtwo$(\tau)$ is sketched in Fig.~\ref{fig_HBT}. A counter is started when a photon is detected on A, and run until another photon is detected on B, which causes the counter to stop. With time accumulating, an histogram of events can be gradually built as function of the duration during which the counter ran. By counting the number of events as a function of the time difference $(\tau)$, we measure \gtwo$(\tau)$. Typical results are illustrated in Fig.~\ref{fig_g2}. In the case of a intensity interferometer, the beam splitter is replaced by the spatial extent of the light source. 

Quantum optics experiments have shown that \gtwo$(0)$ can be smaller than unity, which is the signature of non-classical light. To compare such result to the classical light fluctuations, lets adopt a quantum description in terms of photons. Intensity fluctuations can be understood as "bunches" of photons. In other words, in chaotic light photons tends to arrive in groups because the collection of individual photon emission is randomly disturbed (by, for instance, collisions between atoms in the emitting plasma). These bunches are characterized by a distribution of arrival times that is super poissonian, i.e. with a statistical variance following: $\sigma^2(N) = \bar{N}+\bar{N}^2.$

Similarly, in the section of the light beam of an ideal monochromatic polarized laser, the arrival time of a photon will not be correlated whatsoever with that of another one, \gtwo$(\tau)$ remains equal to unity whatever the value of $\tau$, and the variance is perfectly poissonian: $\sigma^2(N) = \bar{N}$. From a classical point of view, the intensity of such laser has no fluctuations at all, and an infinite coherence time. 

The case of \gtwo$(0) < 1$ is often called anti-bunching. This cannot be interpreted in a classical sense (the light would have a coherence time larger than that of a laser, i.e. larger than infinity) and need a quantum description. To be exact, anti-bunching and sub-poissonian statistics are not exactly the same \citep[see e.g.][]{Singh-1983,Zou-Mandel-1990}, even if they tend to be satisfied simultaneously. Formally, anti-bunched light is light whose degree of second-order coherence satisfies \gtwo$(\tau) > $ \gtwo$(0)$, while light with sub-Poissonian statistics satisfies \gtwo$(0) < 1$ \citep[see also][Chap.~6.5]{Loudon-2000}.

Photon anti-bunching is only one aspect of the quantum properties of light among many others: squeezed states, intricated photons, slow light and so on. It is however the first observable that could be used in astrophysics, requiring only a large photon collector, and ultra-fast and efficient detectors. It is out of the scope of this paper to see if other properties can be studied in practice in an astrophysical context. At this stage, we focus on light statistics and anti-bunching, which still represents truly an unknown territory in optical observations of astrophysical sources.

\section{The Intensity Interferometer: fundamental limitations and new detectors}

It has been presented in other places some considerations about using present and future \v{C}erenkov telescope arrays to build a modern Intensity Interferometer \citep[see in particular][]{LeBohec-Holder-2005,Ofir-Ribak-2006b,deWit-etal-2008b}. It is already known that one of the main issues with II is its much lower sensitivity compared to an amplitude interferometer, since a second-order effect is measured. On the other hand, it is often believed that by improving the various parts of the original II experiment and multiplying the number of baselines/telescopes, one can build a modern interferometer that could open the window towards new scientific results. Although it may certainly be true in practice, we show below that there are at least two fundamental limitations that do not exists with a phase interferometer and that must be kept in mind when looking for science applications of an II.

\subsection{It works for chaotic light only}

We have exposed above the reasons why an intensity interferometer ultimately relies on the chaotic nature of the light produced by the source to make measurements of stellar radii. This hypothesis of chaotic light is certainly verified in the case of thermal light, where the emission processes are dominated by random collisions and Doppler broadening. Hence, a modern intensity interferometer will remain (only) an imaging device of thermal sources. 

Interestingly, second-order coherence has been looked for in radio masers, but no deviation from a super-poissonian statistics has been found \citep{Evans-etal-1972,Moran-1981}. The only interpretation proposed is that even though a maser produce locally coherent light, an astrophysical object called "maser" is made of a collection of individual coherent sources that have random motions relative to each other.

This example illustrate the obvious fact that the interpretation of photon statistics in an astrophysical observation is not necessarily straightforward and requires some a priori knowledge about the source.

\subsection{The hotter, the better}

The second fundamental limitation of an II, to be placed on top of the previous one, is that the amount of intensity fluctuations, i.e. the "signal" used to measure stellar radii, decreases rapidly with the star's surface temperature. In other words, the hotter the better. For a thermal source, the photon statistics follows the Bose-Einstein distribution, whose variance writes:
\begin{equation}
\sigma^2(n) = \bar{n} + \bar{n}^2
\end{equation}
The first term, called the shot noise, is originating from the discrete nature of light (photons) while the second, called wave noise, truly represent the fluctuations of the energy of the electromagnetic radiation. Hence this latter term is the "signal". However, its contribution for the total noise is rather small at optical frequencies.

Following \citet[][Equs. 9.3-8 and 9.5-19]{Goodman-1985}, assuming that individual count-products at the output of the device for a single counting interval are independent from one interval to another, the Signal-to-Noise writes for a blackbody:
\begin{equation}
\left( \frac{S}{N} \right)_{th} = V^2 \sqrt{\frac{T_{\textrm{\scriptsize exp}}}{\tau}} \delta
\end{equation}
where $V$ is the first-order visibility, $T_{\textrm{\scriptsize exp}}$ the exposure time, $\tau$ the shortest integration interval of the detector, and $\delta$ the degeneracy parameter. The degeneracy parameter represents the average number of counts per mode, or per single coherence interval of the incident radiation. In other words, when $\delta \ll 1$, shot noise dominates, while $\delta \gg 1$, the contribution of the intensity fluctuations to the statistics are dominant. For a blackbody:
\begin{equation}
\delta = \frac{1}{\exp(h\nu/k_B T) - 1}
\end{equation}
where $T$ is the source's temperature \citep[see][Chap. 9.5]{Goodman-1985}. We note that wave noise is by far the dominant source of noise in the radio-frequency domain.

\begin{figure}
	\centering
	\includegraphics[width=0.9\linewidth]{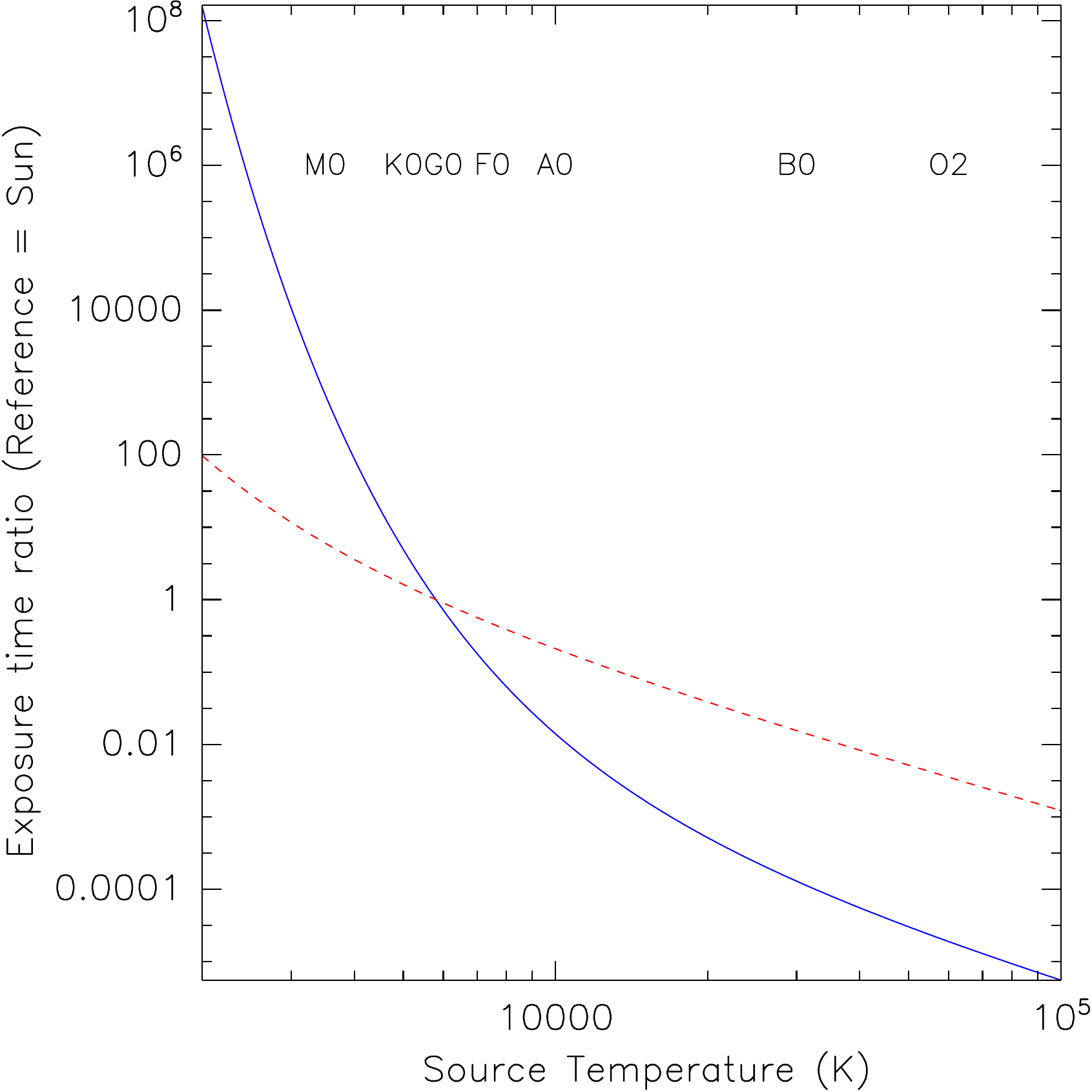}
	\caption{Exposure time required to observe a given S/N ratio with a given visibility and detector integration interval, as a function of the stellar temperature, normalized to solar values. The blue solid curve shows the case for visible light ($\lambda = 0.55\mu$m), while the red dashed curve is for the infrared K-band ($\lambda = 2.5\mu$m). The approximate position of spectral types are also shown.}
	\label{fig_SN}
\end{figure}

For a given Signal-to-Noise ratio, visibility and integration interval, we have computed the exposure times as a function of the star's temperature, and normalized it by solar values. The result is shown in Fig.~\ref{fig_SN}. One can clearly see that, compared to the observation of a solar analog, cooler stars rapidly require much longer exposure times, especially at short wavelengths while it is not so critical when observing at infrared wavelengths. This limitation illustrate the fact that Hanbury Brown \& Twiss observed mostly hot stars, and it must be taken into account when looking for scientific application of a modern II.

\subsection{New photon-counting HgCdTe Avalanche Photodiodes and the quantum limit}

The intensity interferometry is the measurement of a second-order effect and as such, is poorly sensitive. This limitation explains the need for large light collectors, and therefore the concomitant interest for this technique with the future advent of very large optical telescopes. On the other hand, another component not less critical are the detectors. More efficient and rapid detectors relieve slightly the requirements on mirror's size although they increase the requirement of isochronicity of the mirrors. As matter of fact, they are of central importance when looking for light statistics, since the quantum efficiency measures the fidelity of this statistics. Using a semi-classical approach for photodetection, one can show \citep[e.g.][]{Fox-2006} that the observed variance of the statistics of photon arrival times writes:
\begin{equation}
\sigma^2(N) = \eta^2 \sigma^2(n) + \eta (1-\eta)\bar{n}
\end{equation}
where the quantum efficiency $\eta$ can be defined as the ratio of observed mean photon number $\bar{N}$ over that of the true photon number $\bar{n}$ reaching the detector. The above equation shows that a high quantum efficiency is therefore of central importance to have $\sigma^2(N)$ as close as possible to $\sigma^2(n)$.

The other obvious parameter to consider is the smallest time interval accessible by the detector, or inversely its bandwidth. It can be shown that, for instance in the case of chaotic light, the {\it observed} photon statistics variance writes:
\begin{equation}
\sigma^2(n) = \bar{n} + \bar{n}^2 \left(\frac{\tau_c}{\tau}\right)
\end{equation}
where $\tau_c$ is the coherence time of the light, and $\tau$ the exposure time of an individual measurement. If $\tau \gg \tau_c$, the variance reduces to a poissonian one. Interestingly, the above equation is also valid in the case of the NSII, where $\tau_c$ is now the coherence time of the light fluctuations (which is much longer than the coherence time of the thermal light of the observed stars, since $\tau_c \propto 1/(\omega_1-\omega_2)$, see above). 

Detectors considered by the time of QuantEYE prospective work \citep[a concept of photon-counting instrument for the E-ELT, see][and below]{Dravins-etal-2005} were limited to the nanosecond accuracy. However, new perspectives could be opened by a novel type of polarized HgCdTe Avalanche Photodiodes developed at the Commissariat \`a l'Energie Atomique (CEA) in Grenoble \citep{Rothman-etal-2008}. It appears that these diodes could provide for a high gain, a low dark current and a very high quantum efficiency (above 95\%, J. Rothman, private communication) a way to detect single photons in the optical-near infrared domain with an accuracy of time tagging better than 100 picoseconds.

\begin{figure}
	\centering
	\includegraphics[width=0.9\linewidth]{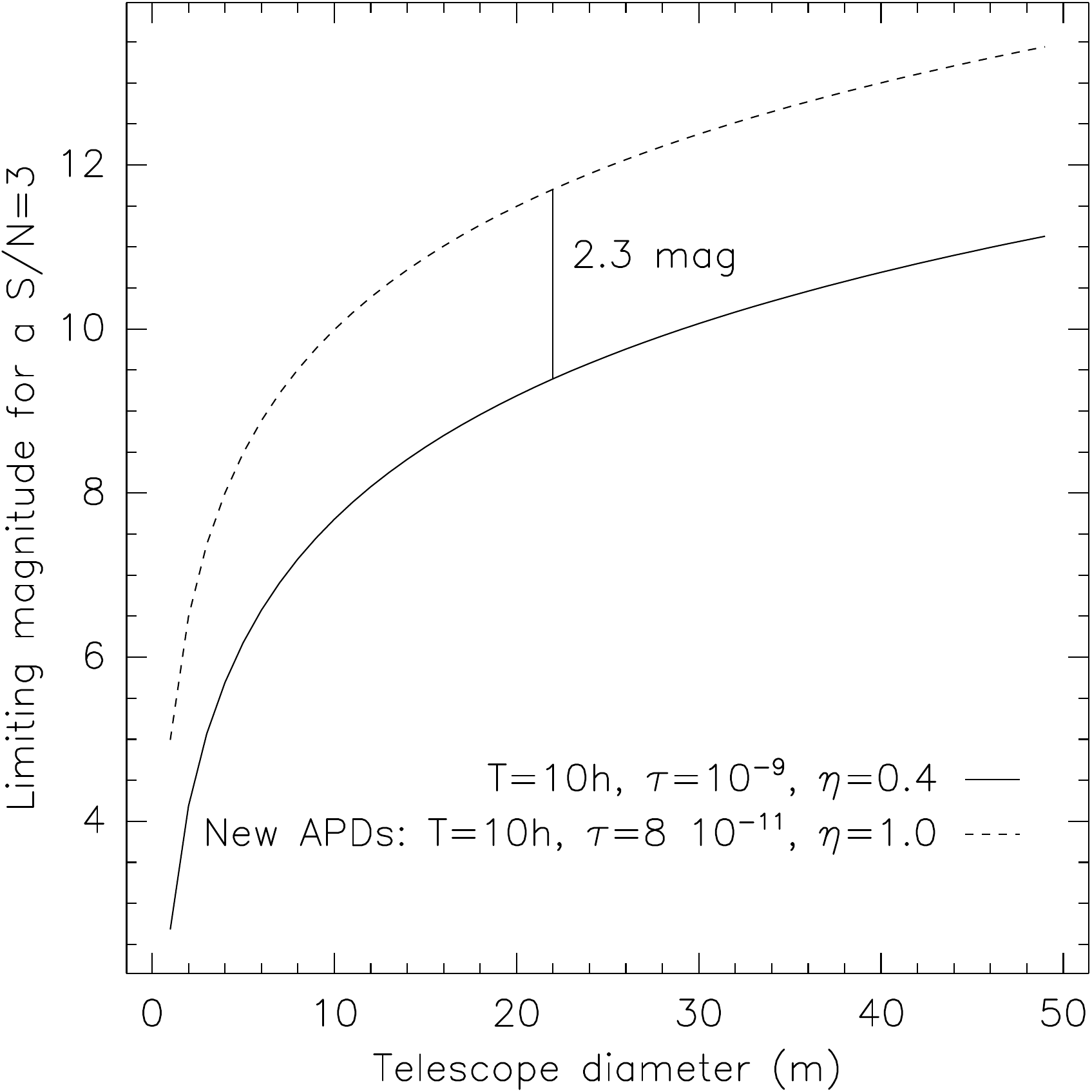}
	\caption{Impact of the new APDs on the limiting magnitude of an II. The comparison is made with \citet{LeBohec-Holder-2005}, using a minimal S/N of 3, and a 10 hours night integration, an overall reflectivity of $R=0.6$, which gives a limiting magnitude of about 8 for a telescope area of $A = 100 \; m^2$ (i.e. d = 11.3 m). The gain with new APDs, with an assumed quantum efficiency of 95\% and a time tagging accuracy of 80 picoseconds is about 2.3 magnitudes.}
	\label{fig_SN_APD_1}
\end{figure}

Fundamentally speaking, one can define an observational "photon-counting quantum limit" by using Heisenberg's uncertainty principle:
\begin{equation}
\Delta t \; \Delta E \geq \hbar
\end{equation}
Using the expected accuracy of the detectors above (80 picoseconds), and a central wavelength of $\lambda \sim 6000$ \AA, the limit is achieved with a spectral resolution of $R \equiv {\lambda}/{\Delta \lambda} \sim 40\,000$. It means that to reach the limit, these detectors could in principle be placed behind an echelle grating in the optical domain similar to what is installed in UVES, the UV-Visual Echelle Spectrograph of ESO's Very Large Telescope.

One can evaluate the impact of such detectors by using a Signal-to-Noise ratio formula of \citet[][see Equ. 4.30, p50.]{HanburyBrown-1974} taking into account the telescope size and the overall quantum efficiency of the system:
\begin{equation}
\left(\frac{S}{N}\right)_{RMS} = A \; \eta \; R \; n \; V^2 \sqrt{\frac{T}{2\tau}}
\end{equation}
where $A$ is the telescope area in square meters, $\eta$ the photodetector quantum efficiency, $R$ the overall optics reflectivity, $n$ the source spectral density in photons per unit optical bandwidth, per unit area and per unit time. $V$ is the visibility, $\tau$ the detector and electronic bandwidth, and $T$ the total integration time in seconds. For the new detectors, we assume a quantum efficiency of 0.95 and an inverse bandwidth of 80 picoseconds. The comparison with typical detectors expected in \v{C}enrenkov intensity interferometers by \citet{LeBohec-Holder-2005} with $\eta = 0.4$ and a bandwidth of GHz are shown in Fig.~\ref{fig_SN_APD_1}. One can see that these new APDs would increase the limiting magnitude by about 2.3.

We emphasize that these estimations are somehow optimistic. \citet{HanburyBrown-1974} observed true S/N systematically lower by 25\% compared to his theoretical calculations given by the above formula. He gave no explanations, but identified the most probable cause of this discrepancy as the cumulative result of a number of small errors in various parameters: overall loss of the optical system, the excess noise and atmospheric effects.

\section{New scientific questions}

A revival of the intensity interferometry is interesting as it may create, among other things, new means to perform high-resolution angular imaging \citep{deWit-etal-2008b}. However, to our opinion, this regain of interest could be broader, and should encompass the perspectives already outlined in the prospective work of the photon-counting instrument QuantEYE \citep[]{Dravins-etal-2005} proposed for the E-ELT \citep[at that time the ELT was called OWL and was expected to have a 100m primary mirror]{DOdorico-2005}, to which the reader is referred for a broad review on observational high-speed astrophysics. In that context, one must explore the new possibilities offered by true photon-counting devices. We note that some experiments already started at the Asagio observatory \citep[AquaEYE;][]{Barbieri-etal-2007,Naletto-etal-2007}. 

With truly new possibilities, we think it is important to explore new questions, that do not simply go slightly beyond what is accessible by the current (or even foreseen) "classical" observational techniques. In this section we attempt at contributing to these efforts by presenting why microquasars in our galaxy, and possibly their extragalactic equivalent, the quasars, could represent an excellent first target in the optical/near-infrared where to observe non-thermal light, and test the use of \gtwo\ in astrophysical sources.

We note that Hanbury Brown judged perfectly unlikely that the coincidence-counting interferometry was of any interest for astrophysics. The reason he gives (end of Chap. 4) is however essentially technical and related to the fact that bandwidth must be limited in order to not saturate the detector and permit the counting of every single photon. The example he took is however that of a zero-th magnitude star.

\subsection{Microquasars under the quantum microscope}

Microquasars are short-period X-ray binaries with one of its component being a stellar-mass black-hole. These objects are the closest relativistic objects to us. They sometimes show superluminal jets \citep[e.g.][]{Mirabel-Rodriguez-1994}, and produce copious amount of X-rays. Powerful and self-collimated jets are produced at the inner regions of the accretion disk \citep[e.g.][]{Ferreira-etal-2006}. Microquasars share the same physics as quasars and probably gamma-ray bursts (since they share the same physical ingredients: a black-hole, an accretion disk and jets); they must therefore be understood in the same framework \citep{Mirabel-2004}. Microquasars are nowadays studied mostly in X-rays because it is where the physics of the closest regions around the black-hole can be probed. 

Interestingly, the jets in microquasars are supersonic, which means that the particles emitting synchrotron radiation are moving faster than the local perturbations, possibly avoiding the problem observed in radio masers. Moreover, the jets are being produced very close to the central black-hole. Therefore, jets could possibly used to probe the region around the black-hole similarly to what is made nowadays: the accretion disk and its oscillations \citep[called Quasi-Periodic Oscillation - QPOs, see for instance][]{vanderKlis-2005} can be used to estimate black-hole spin \citep{Remillard-McClintock-2006}.

\begin{figure}
\centering
\includegraphics[width=0.5\linewidth]{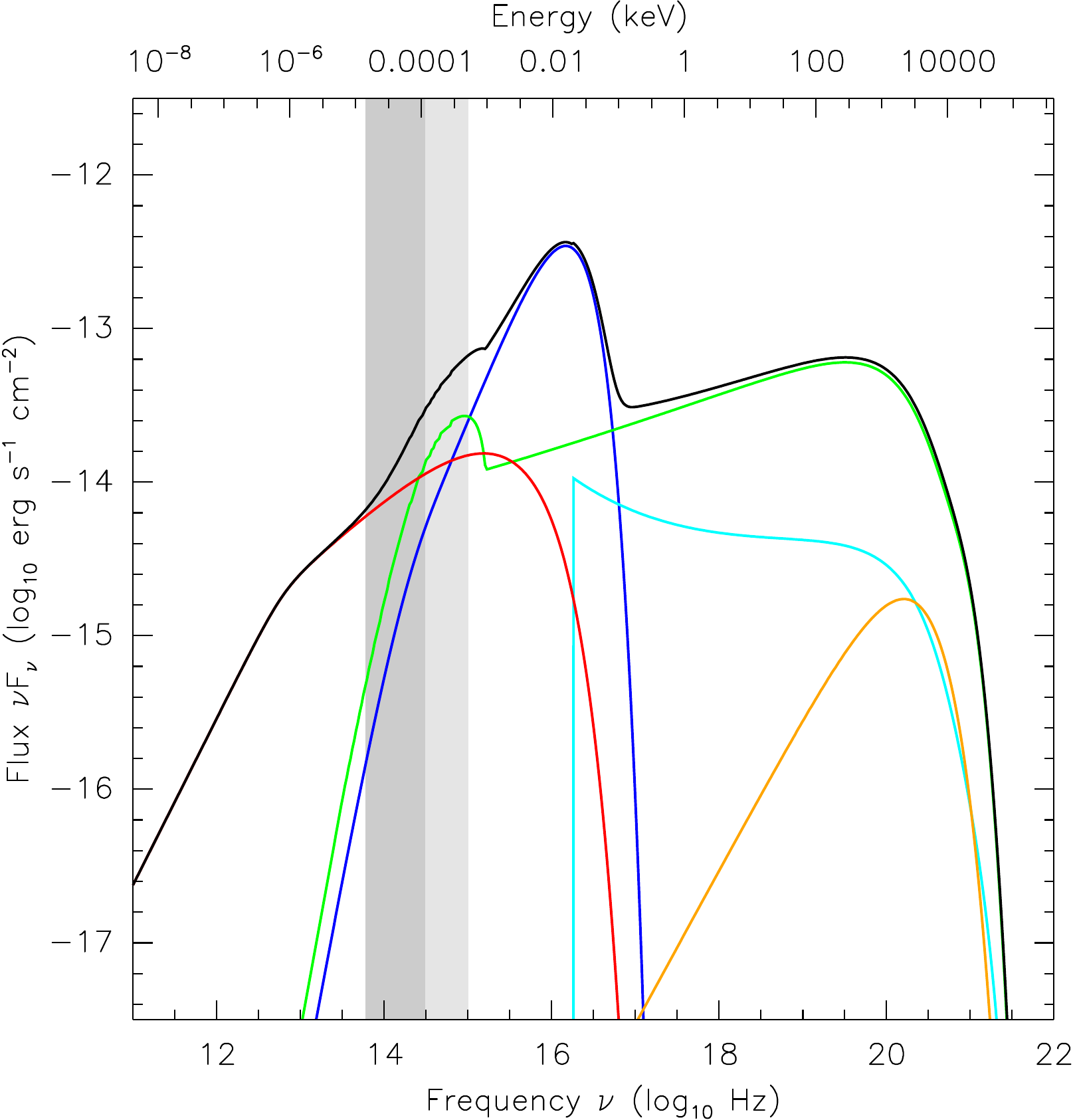}\includegraphics[width=0.5\linewidth]{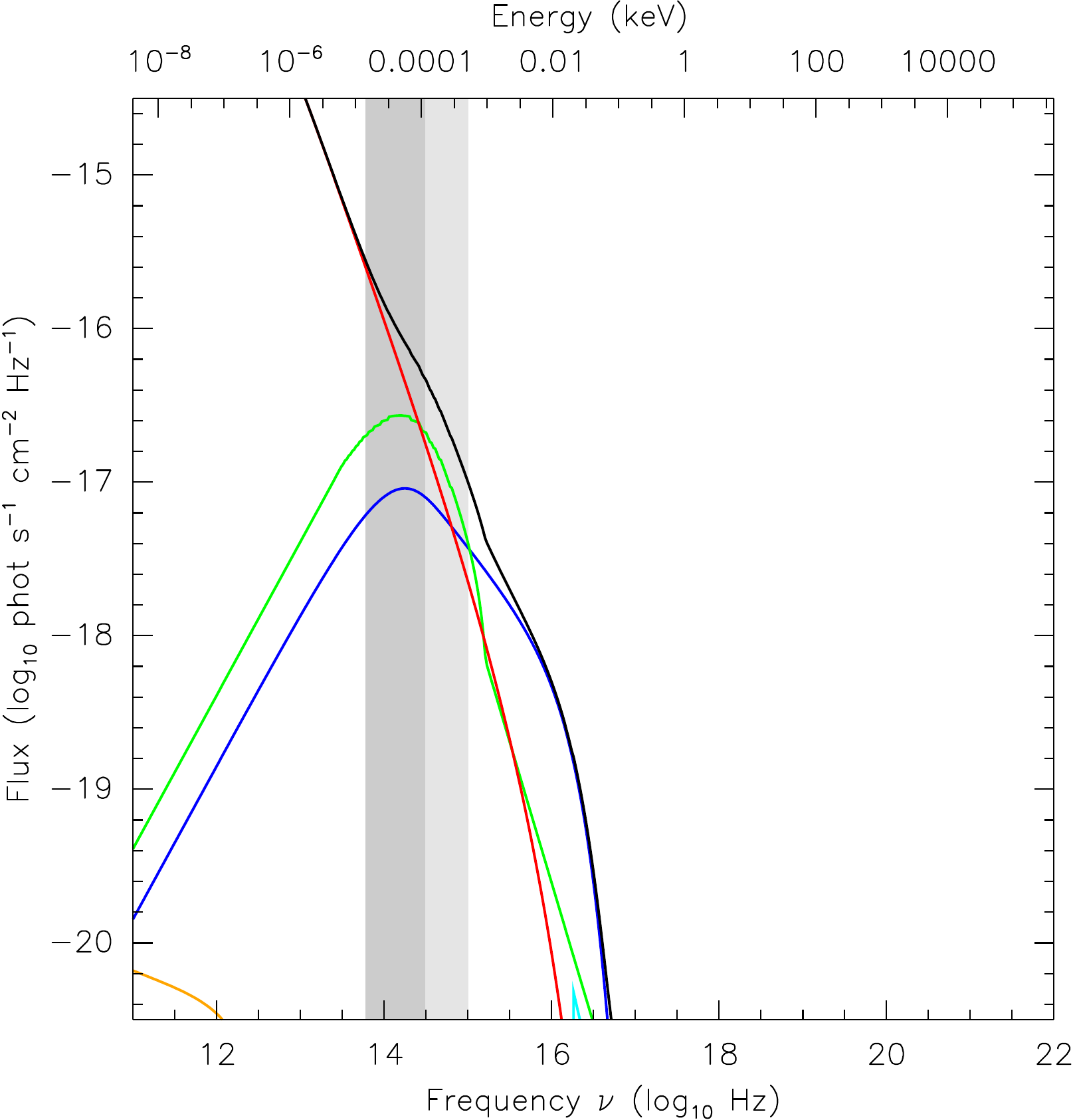}
\caption{Theoretical Spectral Energy Distributions of a microquasar as a function of the frequency $\nu$. In red the synchrotron radiation of the jet, in blue the multi-temperature black-body radiation of the external disk, in orange the Bremsstrahlung, in green the Comptonized synchrotron emission from the inner disk, and in cyan the Comptonization of the external disk \citep[see e.g.][]{Foellmi-etal-2008a,Foellmi-etal-2008b} + Foellmi et al. 2009). Left: flux ($\nu F_{\nu}$) expressed in flux units . Right: flux expressed in photon count rate. Distance of the object is 10~kpc. Optical and NIR regions of the spectrum are shown by gray areas.}
\label{fig_seds}
\end{figure}

The optical/near-infrared region of the spectrum is well suited for the study of the interface between the jets and the accretion disk around the black-hole. Fig.~\ref{fig_seds} illustrate a typical theoretical spectral energy distribution (SED) of a microquasar with a black-hole mass of 10$M_{\odot}$ at a distance of 10~kpc \citep[see e.g.][]{Foellmi-etal-2008a}, (Foellmi et al. 2009, MNRAS, in prep). The left panel shows the observed flux of the various radiating components in erg~s$^{-1}$~cm$^{-2}$. The most central regions of the accretion disk around the black hole produce the strong X-ray emission. However, the exact same SED expressed in terms of photons (right panel) shows a completely different situation. In the far-UV and X-ray domain, the number of photons is simply much too small to be used for statistical studies in a reasonable amount of time. 

Finally, let us mention an example of another quantum property of the light produced, in theory, nearby a black-hole: the Unruh process \citep[a cinematic version of the Hawking radiation,][]{Unruh-1976} is expected to produce pairs of photons with maximally entangled polarizations \citep{Schutzhold-etal-2006}.

\section{Summary and conclusions}

In this research note we have presented the mathematical basics about the first and the second order correlation functions, and discussed them in the context of astronomical observations. 

It appears rather clear that detailed physical and quantitative predictions of quantum phenomena in astrophysical sources accessible with a photon-counting devices are still lacking. In the context of the development of large \v{C}erenkov arrays and future Extremely Large Telescopes, we think that these questions deserve more dedicated studies.

\section*{Acknowledgments}
C.F. thanks E. Le Coarer, J. Rothman, D. Dravins, P. Kern, G. Duvert, D. Mouillet, S. Maret \& A. Chelli for exciting discussions, J.-P. Berger for providing a very useful reference, K. Schuster for encouragements. C.F. also acknowledges support from the Swiss National Science Foundation (grant PA002--115328).



\begin{thebibliography}{42}
\expandafter\ifx\csname natexlab\endcsname\relax\def\natexlab#1{#1}\fi

\bibitem[{Alexander(2003)}]{Alexander-2003}
Alexander, G. 2003, Rep. Prog. Phys., 66, 481

\bibitem[{Barbieri {et~al.}(2007)Barbieri, Naletto, Occhipinti, Tamburini,
  Giro, D'Onofrio, Sain, \& Zaccariotto}]{Barbieri-etal-2007}
Barbieri, C., Naletto, G., Occhipinti, T., {et~al.} 2007, Mem. S.
  A. It. Supp., 11, 190

\bibitem[{Baym(1998)}]{Baym-1998}
Baym, G. 1998, Acta Physica Polonica B, 29, 1839

\bibitem[{Borra(2008)}]{Borra-2008}
Borra, E.~F. 2008, MNRAS, 389, 364

\bibitem[{Brown(1974)}]{HanburyBrown-1974}
Brown, R.~H. 1974, New York, Halsted Press

\bibitem[{Brown \& Twiss(1957)}]{HanburyBrown-Twiss-1957a}
Brown, R.~H. \& Twiss, R.~Q. 1957, Proc. of the Royal Society of London.
  Series A, 242, 300

\bibitem[{Brown \& Twiss(1958{\natexlab{a}})}]{HanburyBrown-Twiss-1958a}
---. 1958{\natexlab{a}}, Proc. of the Royal Society of London. Series A,
  243, 291

\bibitem[{Brown \& Twiss(1958{\natexlab{b}})}]{HanburyBrown-Twiss-1958b}
---. 1958{\natexlab{b}}, Proc. of the Royal Society of London. Series A,
  248, 199

\bibitem[{Brown \& Twiss(1958{\natexlab{c}})}]{HanburyBrown-Twiss-1958c}
---. 1958{\natexlab{c}}, Proc. of the Royal Society of London. Series A,
  248, 222

\bibitem[{de~Wit {et~al.}(2008{\natexlab{a}})de~Wit, Bohec, Hinton, White,
  Daniel, \& Holder}]{deWit-etal-2008a}
de~Wit, W.~J., Bohec, S.~L., Hinton, J.~A., {et~al.} 2008{\natexlab{a}}, High
  Time Resolution Astrophysics: The Universe at Sub-Second Timescales. AIP
  Conf. Proc., 984, 268

\bibitem[{de~Wit {et~al.}(2008{\natexlab{b}})de~Wit, LeBohec, Hinton, White,
  Daniel, \& Holder}]{deWit-etal-2008b}
de~Wit, W.~J., LeBohec, S., Hinton, J.~A., {et~al.} 2008{\natexlab{b}}, Journal
  of Physics: Conference Series, 131, 2050

\bibitem[{D'Odorico(2005)}]{DOdorico-2005}
D'Odorico, S. 2005, The Messenger (ESO), 122, 6, (c) 2005: European Southern
  Observatory

\bibitem[{Dravins(2008)}]{Dravins-2008}
Dravins, D. 2008, High Time Resolution Astrophysics, Astrophysics and Space
	Science Library, Springer, 351, 95

\bibitem[{Dravins {et~al.}(2005)Dravins, Barbieri, Fosbury, Naletto, Nilsson,
  Occhipinti, Tamburini, Uthas, \& Zampieri}]{Dravins-etal-2005}
Dravins, D., Barbieri, C., Fosbury, R. A.~E., {et~al.} 2005,
  astro-ph/0511027

\bibitem[{Evans {et~al.}(1972)Evans, Rydbeck, \& Kollberg}]{Evans-etal-1972}
Evans, N., Rydbeck, O. E.~H., \& Kollberg, E. 1972, Phys. Rev. A, 6, 1643

\bibitem[{Ferreira {et~al.}(2006)Ferreira, Petrucci, Henri, Saug{\'e}, \&
  Pelletier}]{Ferreira-etal-2006}
Ferreira, J., Petrucci, P.-O., Henri, G., Saug{\'e}, L., \& Pelletier, G. 2006,
  A{\&}A, 447, 813

\bibitem[{Foellmi {et~al.}(2008{\natexlab{a}})Foellmi, Petrucci, Ferreira, \&
  Henri}]{Foellmi-etal-2008a}
Foellmi, C., Petrucci, P.~O., Ferreira, J., \& Henri, G. 2008{\natexlab{a}},
  arXiv/0810.108

\bibitem[{Foellmi {et~al.}(2008{\natexlab{b}})Foellmi, Petrucci, Ferreira,
  Henri, \& Boutelier}]{Foellmi-etal-2008b}
Foellmi, C., Petrucci, P.-O., Ferreira, J., Henri, G., \& Boutelier, T.
  2008{\natexlab{b}}, SF2A-2008: Proc. of the Annual meeting of the
  French Society of Astronomy and Astrophysics Eds.: C. Charbonnel, 211

\bibitem[{Fox(2006)}]{Fox-2006}
Fox, M. 2006, Quantum Optics: An Introduction, Oxford University Press

\bibitem[{Glauber(2007)}]{Glauber-2007}
Glauber, R.~J. 2007, Quantum Theory of Optical Coherence: Selected Papers and
  Lectures, Wiley-VCH, 639

\bibitem[{Goodman(1985)}]{Goodman-1985}
Goodman, J.~W. 1985, Statistical Optics, Wiley

\bibitem[{Haniff(2007)}]{Haniff-2007a}
Haniff, C. 2007, New Astronomy Reviews, 51, 565

\bibitem[{Holmes \& Belen'kii(2004)}]{Holmes-Belenkii-2004}
Holmes, R.~B. \& Belen'kii, M.~S. 2004, Journal of the Optical Society of
  America A, 21, 697

\bibitem[{Jain \& Ralston(2008)}]{Jain-Ralston-2008}
Jain, P. \& Ralston, J.~P. 2008, A{\&}A, 484, 887

\bibitem[{LeBohec {et~al.}(2008)LeBohec, Barbieri, Wit, Dravins, Feautrier,
  Foellmi, Glindemann, Hall, Holder, Holmes, Kervella, Kieda, Coarer, Lipson,
  Malbet, Morel, Nunez, Ofir, Ribak, Saha, Shoeller, Zhilyaev, \&
  Zinnecker}]{LeBohec-etal-2008}
LeBohec, S., Barbieri, C., Wit, W.-J.~D., {et~al.} 2008, Proc. of the
  SPIE, 7013, 70132E

\bibitem[{LeBohec \& Holder(2005)}]{LeBohec-Holder-2005}
LeBohec, S. \& Holder, J. 2005, 29th International Cosmic Ray Conference,
  astro-ph/057010

\bibitem[{Loudon(2000)}]{Loudon-2000}
Loudon, R. 2000, The Quantum Theory of Light, Oxford University Press

\bibitem[{Mirabel(2004)}]{Mirabel-2004}
Mirabel, I.~F. 2004, Proceedings of the 5th INTEGRAL Workshop on the INTEGRAL
  Universe (ESA SP-552). 16-20 February 2004, 552, 175

\bibitem[{Mirabel \& Rodriguez(1994)}]{Mirabel-Rodriguez-1994}
Mirabel, I.~F. \& Rodriguez, L.~F. 1994, Nature, 371, 46

\bibitem[{Moran(1981)}]{Moran-1981}
Moran, J.~M. 1981, BAAS, 13, 508

\bibitem[{Naletto {et~al.}(2007)Naletto, Barbieri, Occhipinti, Tamburini,
  Billotta, Cocuzza, \& Dravins}]{Naletto-etal-2007}
Naletto, G., Barbieri, C., Occhipinti, T., {et~al.} 2007, Proc. of the
  SPIE, 6583, 9

\bibitem[{Ofir \& Ribak(2006{\natexlab{a}})}]{Ofir-Ribak-2006a}
Ofir, A. \& Ribak, E.~N. 2006{\natexlab{a}}, MNRAS, 368, 1646

\bibitem[{Ofir \& Ribak(2006{\natexlab{b}})}]{Ofir-Ribak-2006b}
---. 2006{\natexlab{b}}, MNRAS, 368, 1652

\bibitem[{Remillard \& McClintock(2006)}]{Remillard-McClintock-2006}
Remillard, R.~A. \& McClintock, J.~E. 2006, ARA{\&}A, 44, 49

\bibitem[{Rothman {et~al.}(2008)Rothman, Perrais, Ballet, Mollard, Gout, \&
  Chamonal}]{Rothman-etal-2008}
Rothman, J., Perrais, G., Ballet, P., {et~al.} 2008, Journal of Electronic
  Materials, 37, 1303

\bibitem[{Schawlow \& Townes(1958)}]{Schawlow-Townes-1958}
Schawlow, A.~L. \& Townes, C.~H. 1958, Phys. Rev., 112, 1940

\bibitem[{Sch{\"u}tzhold {et~al.}(2006)Sch{\"u}tzhold, Schaller, \&
  Habs}]{Schutzhold-etal-2006}
Sch{\"u}tzhold, R., Schaller, G., \& Habs, D. 2006, Phys. Rev. Let., 97

\bibitem[{Singh(1983)}]{Singh-1983}
Singh, S. 1983, Optics Communications, 44, 254

\bibitem[{Unruh(1976)}]{Unruh-1976}
Unruh, W.~G. 1976, Phys. Rev. D, 14, 870

\bibitem[{van~der Klis(2005)}]{vanderKlis-2005}
van~der Klis, M. 2005, Astronomische Nachrichten, 326, 798

\bibitem[{Zhilyaev(2008)}]{Zhilyaev-2008}
Zhilyaev, B.~E. 2008, astro-ph/0806.0191

\bibitem[{Zou \& Mandel(1990)}]{Zou-Mandel-1990}
Zou, X.~T. \& Mandel, L. 1990, Phys. Rev. A, 41, 475

\end{thebibliography}
\end{document}